\documentclass[%
 aip,
 amsmath,amssymb,
 reprint,%
]{revtex4-1}

\usepackage{graphicx}
\usepackage{dcolumn}
\usepackage{bm}

\usepackage[utf8]{inputenc}
\usepackage[T1]{fontenc}
\usepackage{mathptmx}
\usepackage{etoolbox}

\usepackage{amsmath}
\usepackage[colorlinks=true, linkcolor=blue, urlcolor=blue, citecolor=blue]{hyperref} 
\usepackage{url}
\usepackage{subcaption}
\usepackage{orcidlink}

\usepackage{placeins}

\makeatletter
\def\@email#1#2{%
 \endgroup
 \patchcmd{\titleblock@produce}
  {\frontmatter@RRAPformat}
  {\frontmatter@RRAPformat{\produce@RRAP{*#1\href{mailto:#2}{#2}}}\frontmatter@RRAPformat}
  {}{}
}%
\makeatother
\begin{document}

\preprint{AIP/123-QED}

\title{Low order Carleman simulation of steady state flows at mild Reynolds numbers}
\author{Luca Cappelli~\orcidlink{0009-0009-1169-8380}}
\email{luca.cappelli@iit.it}
\affiliation{ 
Fondazione Istituto Italiano di Tecnologia, Center for Life Nano-Neuroscience at la Sapienza, Viale Regina Elena 291, 00161 Roma, Italy
}

\author{Massimo Bernaschi}
\affiliation{Istituto per le Applicazioni del Calcolo ``Mauro Picone", Italian National Research Council}

\author{Sauro Succi\orcidlink{0000-0002-3070-3079}}%
\affiliation{ 
Fondazione Istituto Italiano di Tecnologia, Center for Life Nano-Neuroscience at la Sapienza, Viale Regina Elena 291, 00161 Roma, Italy
}%

\date{\today}




\begin{abstract}
It is shown that the second-order Carleman truncation (C2) captures the steady state of 
the decaying forced logistic with second-order accuracy in the forcing parameter, provided such 
forcing remains below a critical value of $1/2$. 
We then investigate the extension of these results to spatially extended systems by 
means of Carleman Lattice Boltzmann simulations of a two-dimensional periodically forced flow. 
It is found that C2 correctly reproduces the low-Reynolds-number 
(between 1 and 10) quasi-linear dynamics, well above the critical value of 
the forced logistic equation. However, C2 fails to capture more complex steady states characterised
by persistent vortical structures at Reynolds numbers around 20. 
Interestingly, the divergence observed in the latter case shows qualitative similarities  with 
the runaway behaviour of the C2 linearization of the growing logistics, suggesting  that 
suitable closures and/or change of variables may improve the convergence of the C2 procedure.
\end{abstract}

\maketitle

\section{Introduction}

Carleman linearization provides an effective way of transforming a finite-dimensional
nonlinear system into an infinite-dimensional linear one~\cite{Carleman}.
The technique is long known and it has found applications mostly for the
analytical and computational study of low-dimensional nonlinear dynamical
systems~\cite{HERNANDEZ2004745, TsiligiannisLyberatos1989, Kowalski1991, HASEMIAN2015, SVORONOS1994, STEEB1980, HELLEMAN1974, Munoz_2025}.
In recent years, it has witnessed  a major surge of interest in connection with
the development of quantum algorithms for classical systems, most notably fluid flows~\cite{Itani2022, itani_24, Sanavio_2024, forets2017, Conde2025, Cappelli26, Amini2025, Zamora2026, wiebe25, meng_clb, Liu2021, Mezzacapo2015, joseph25, Setty_2026, Turro_25, demirdjan, Wu_2025, Lewis2024, katsushiro}.
The basic idea behind the Carleman procedure is to trade non-linearity for higher 
dimensions (and non-locality). For extended systems, governed by nonlinear classical
field theories, this turns into an exponential dimensional growth: the $K$-th order
Carleman linearization of a $d$-dimensional nonlinear problem is equivalent to
a linear problem in $Kd$ dimensions. This shows that the technique becomes
rapidly unviable on classical computers beyond the lowest-order truncations, say $K\geq4$.
Quantum computers, on the other hand, could in principle handle much higher 
truncation orders thanks to the exponential power of qubits versus classical bits.
However, simulating the Carleman system on quantum computers raises a formidable 
challenge due to the memory constraints imposed by dimensionality as well
as unitarity constraints~\cite{Sanavio_25_PoF, psiq_26, NielsenChuang2000}. 

The above constraints provides a strong motivation to explore the 
potential of low-order Carleman linearization, with special focus on the
time-asymptotic properties of extended nonlinear dissipative systems.  

In this paper, we present a simple strategy to improve the ability of low-order Carleman 
approximants to capture the time-asymptotic solution of the forced logistic equation. 
Specifically, we analyze the use of closure rather than truncation, and investigate how the 
divergence observed in the growing logistic case can be addressed.
\\
The analysis is then extended to nonlinear spatially extended dynamical systems, focusing on the simulation of two-dimensional homogeneous incompressible flows at low and mild Reynolds numbers within 
the lattice Boltzmann framework. Through numerical simulations on a $64\times64$ 
lattice, performed with a optimized multi-GPU code, 
we empirically identify the regime in which C2 is able to reproduce 
both the system dynamics and its stationary state. In this context, the analysis of 
the logistic equation, despite proving over-restrictive, provides 
nonetheless useful insights into the limitations of low-order Carleman 
approximations for fluid flows and helps identify possible strategies to overcome them.


\section{Decaying logistic equation with external forcing}

Let us consider the forced logistic equation in (0+1) dimensions

\begin{equation}
\label{LOGI0}
\dot x  = -ax + b x^2 + f \,,
\end{equation}
with $a>0$, $b>0$ and $f>0$ constant parameters
and initial condition $x(0)=x_0$.
\\
For the sake of simplicity, it is expedient to rescale
variables as follows $t \to at$, $x \to x/c$, where $c=a/b$ is the 
so-called {\it capacity}.
\\
With this rescaling, the ``reduced'' logistic simplifies to:

\begin{equation}
\label{LOGI1}
\dot x  = - x(1-x) + g^2 \,,
\end{equation}
where the parameter 
\begin{equation}
g^2 = fb/a^2 \,,
\end{equation}
defines the competition between forcing and growth versus decay mechanisms.

The dynamics admits two fixed points:
\begin{equation}
x^*_{\mp} = \frac{1 \mp \gamma}{2} \,,
\end{equation}
with $\gamma = \sqrt{1-4g^2}$.
In the force-free limit $f \to 0$, $x^*_{-} \to 0$ and
$x^*_{+} \to 1$, it recovers the stable and unstable fixed
points of the freely decaying logistic.
Note that $g^2$ is the rescaled value of the linear steady-state $f/a$.

\begin{figure}
\centering
\includegraphics[width=\linewidth]{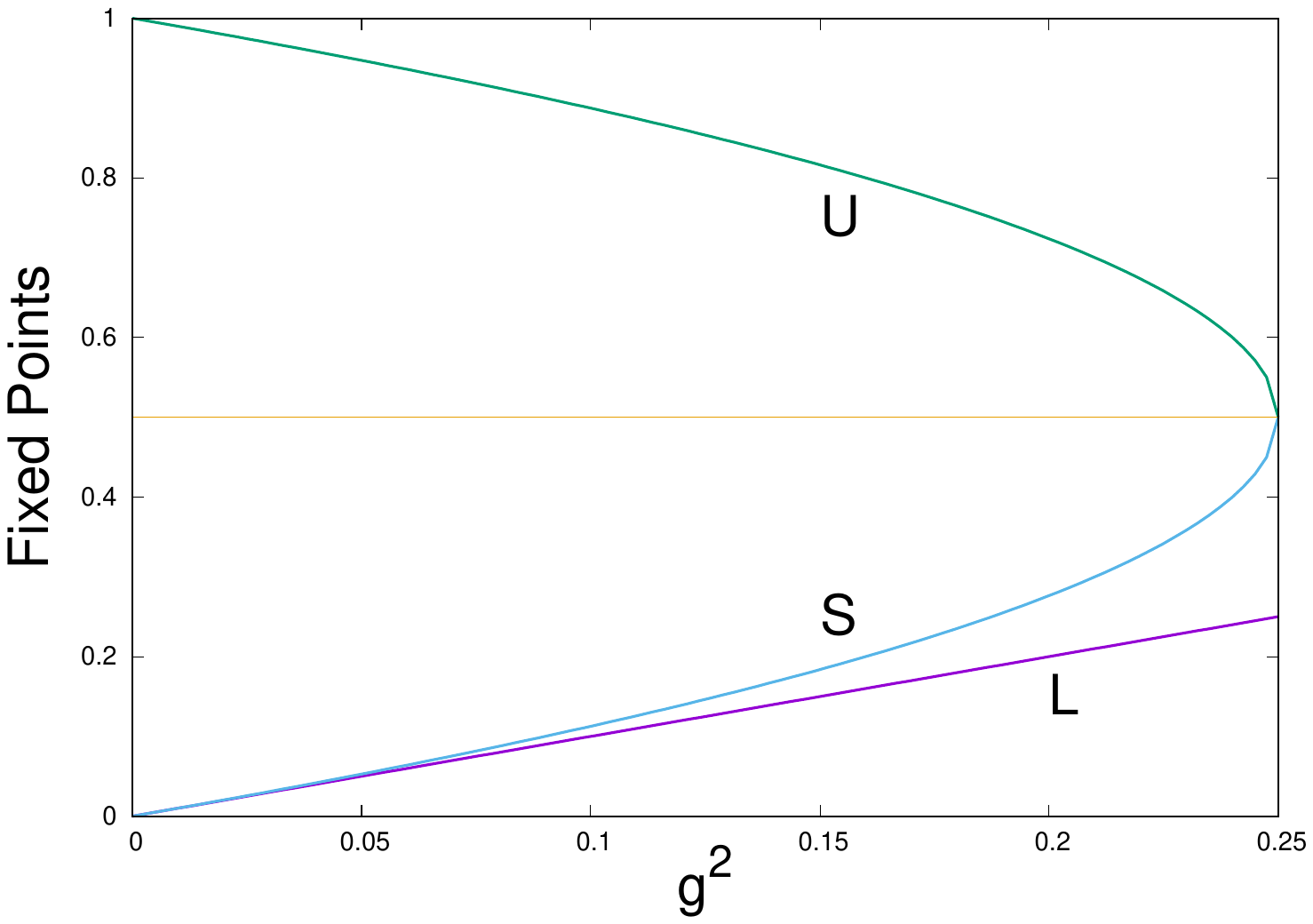}
\caption{The stable state $S \equiv x^*_{-}$, the unstable
one $U \equiv x^*_{+}$ and the linear steady $L$ state $g^2$
as a function of $g^2$.
}
\end{figure}
%
At the critical value $g_c=1/2$, the two fixed points merge. For $g>g_c$, no real-valued steady state exists, and the solutions diverge to infinity in finite time.
\\
The exact solution is shown to be:

\begin{equation}
\label{LOGIE}
x(t)  = \frac{S+U y(t)}{1+y(t)}  \,,
\end{equation}
where $S, U$ denote the stable and unstable fixed points 
and the transformed variable
\begin{equation}
\label{MOBIUS}
y(x) = \frac{x-S}{U-x} \,,
\end{equation}
decays exponentially, $y(t) = y_0 e^{-\gamma t}$, with 
$y_0 = (x_0-S)/(U-x_0)$.

In the long-time limit $\gamma t \gg 1$, $y \to 0$ 
and the solution tends to the stable steady-state 
$x \to S$. In the short time limit $\gamma t \ll 1$,
$y \to y_0$ and it is readily seen that 
$x \to S y_0 + U (1-y_0) \equiv x_0$.
 
The above shows that the (Mobius) transformation 
$x \to y(x)$ linearizes the logistic equation, so
that $e^{-\gamma t}$ serves as a compact expansion parameter to
probe the long-term dynamics of the forced logistic, a property that
connects to the time asymptotics of low-order Carleman linearization. 

In passing, we note that in the force-free case $f \to 0$, we have $g^2 \to 0$, 
$\gamma \to1$, $S \to 0$, $U \to 1$, $y_0 \to x_0/(1-x_0)$ and the exact solution 
reduces to the standard expression $x(t)= x_0 e^{-t} / (1-x_0+x_0 e^{-t})$.

\section{Carleman embedding of the forced logistic}

The Carleman procedure consists of renaming 
$x_1  \equiv x$, $x_2 \equiv x^2$, $x_k \equiv x^k$, 
realizing a linear embedding of the zero-dimensional nonlinear
logistic equation into formal infinite dimensional one. Upon choosing a 
truncation order $K$, the equivalent system is mapped into a $K$-dimensional linear one.

As a result, Eq.~\eqref{LOGI0} takes the form 
\begin{equation}
\label{LOGI1_}
\dot x_1 = - x_1 + x_2 + g^2 \,.
\end{equation}
This is linear but open, hence requiring a dynamic equation for $x_2$.
This is readily obtained by multiplying Eq.~\eqref{LOGI0} by $x$,  
\begin{equation}
\dot x_2 = -2x_2 + 2x_3 + 2 g^2 x_1 \,.
\end{equation}
Iterating $k$ times:
\begin{equation}
\label{CDOT}
\dot x_{k} = -kx_k + kx_{k+1} + kg^2 x_{k-1} + b_k \,,
\end{equation}
where $b_1 = g^2$ and $b_k=0$ for $k>1$.

The formal solution of this system of first order ODE's is:
\begin{equation}
X(t) = e^{Ct} X(0) + (1-e^{Ct}) C^{-1} B \,,
\end{equation}
where $X \equiv [x_1 \dots x_k]$, $B=[b_1 \dots b_k]$
and $C$ is the  tridiagonal Carleman matrix with elements
$C_{k,k-1}= kg^2$, $C_{k,k}=-k$, $C_{k,k+1}=+k$.

In the above we have assumed that $C$ is non-singular and negative-definite
so as to secure stability, as long as $x_0<U$.  
The formal solution becomes operational whenever the full spectrum
of eigenvectors/values of the Carleman matrix $C$ is known, which
is indeed the case for tridiagonal matrices such as those above.

In general, however, an analytical solution is not available
and one usually proceeds by integrating eq. (\ref{CDOT}) numerically 
and closing by truncation, $x_k=0,\;k>K$
or some other linear recurrence based on low-order components.

\section{Carleman representation of the time asymptotics behaviour}

The Carleman-1 (C1) level delivers
\begin{equation}
    \dot x_1 = -x_1 + g^2   \,, 
\end{equation}
whence the C1 steady state $x_1^{ss,1}=g^2$, the linear value.

The Carleman-2 level reads as follows:

\begin{eqnarray}
\dot x_1 = -x_1 + x_2 + g^2 \,,\\
\dot x_2 = -2x_2 + 2x_3 + 2 g^2 x_1 \,.
\end{eqnarray}

At steady state, setting $x_3=0$, we obtain
$x_2 = g^2 x_1$, and replacing into the first equation, we obtain 
\begin{equation}
x_1^{ss,2} = \frac{g^2}{1-g^2} \,.
\end{equation}
Compare now with the stable steady state $S=(1-\sqrt{1-4g^2})/2$.
To first order in $g^2$, we have $S = g^2$ which is exactly the expression
we obtain by expanding $x_1$ to first order in $g^2$.
This shows that the Carleman-2 approximation captures the logistic steady state
to accuracy $O(g^4)$.
As observed in the previous Section, this connects to the fact that Carleman linearization
is directly related to the power series expansion of the exact logistic solution  
in the long-term timescale $\gamma t \gg 1$. 
Clearly, the approximation improves by raising the order of the Carleman truncation.
For instance, a straightforward calculation of the 
Carleman-3 steady state delivers
\begin{equation}
x_1^{ss,3} = \frac{g^2}{1-g^2/(1-g^2)} \,,
\end{equation}
which is accurate to order $g^6$.
Proceeding recursively shows that 
the Carleman approximants to all subsequent orders
correspond to the continued fraction expansion of 
the exact solution $S=(1-\sqrt{1-4g^2})/2$.

\begin{figure}
\centering
\includegraphics[width=.9\linewidth]{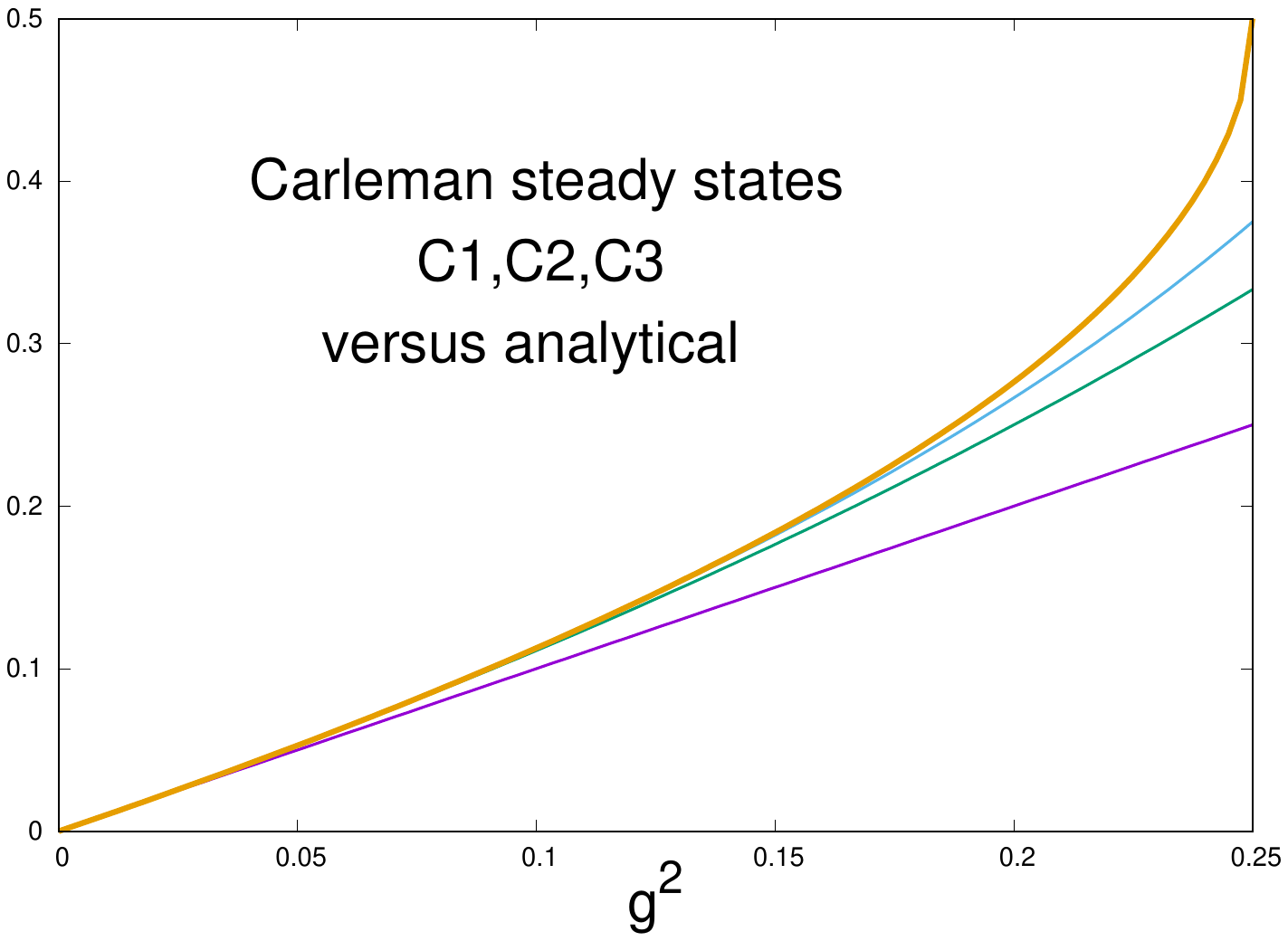}
\caption{The Carleman steady states C1,  C2, C3,
$g^2$, $g^2/(1-g^2)$, $g^2/(1-g^2/(1-g^2))$ versus the analytical
one as a function of $g^2$.
}
\end{figure}

\section{C2 as a false transient technique}

As a numerical example, we integrate the logistic equation for stochastic forcing
to probe the ability of the Carleman-2 approximation to reproduce statistical steady
states as well.

To this purpose, we augment the force term with a random component
\begin{equation}
\eta  =  f_r (2r-1) \,,
\end{equation}
with $r$ a random number uniformly distributed in $[-1:1]$.
The randomness is inserted to inspect whether C2 filters out short-scale fluctuations, a
question that we shall be pursued in more depth in future publications. 

\begin{figure}
\centering
\includegraphics[width=.9\linewidth]{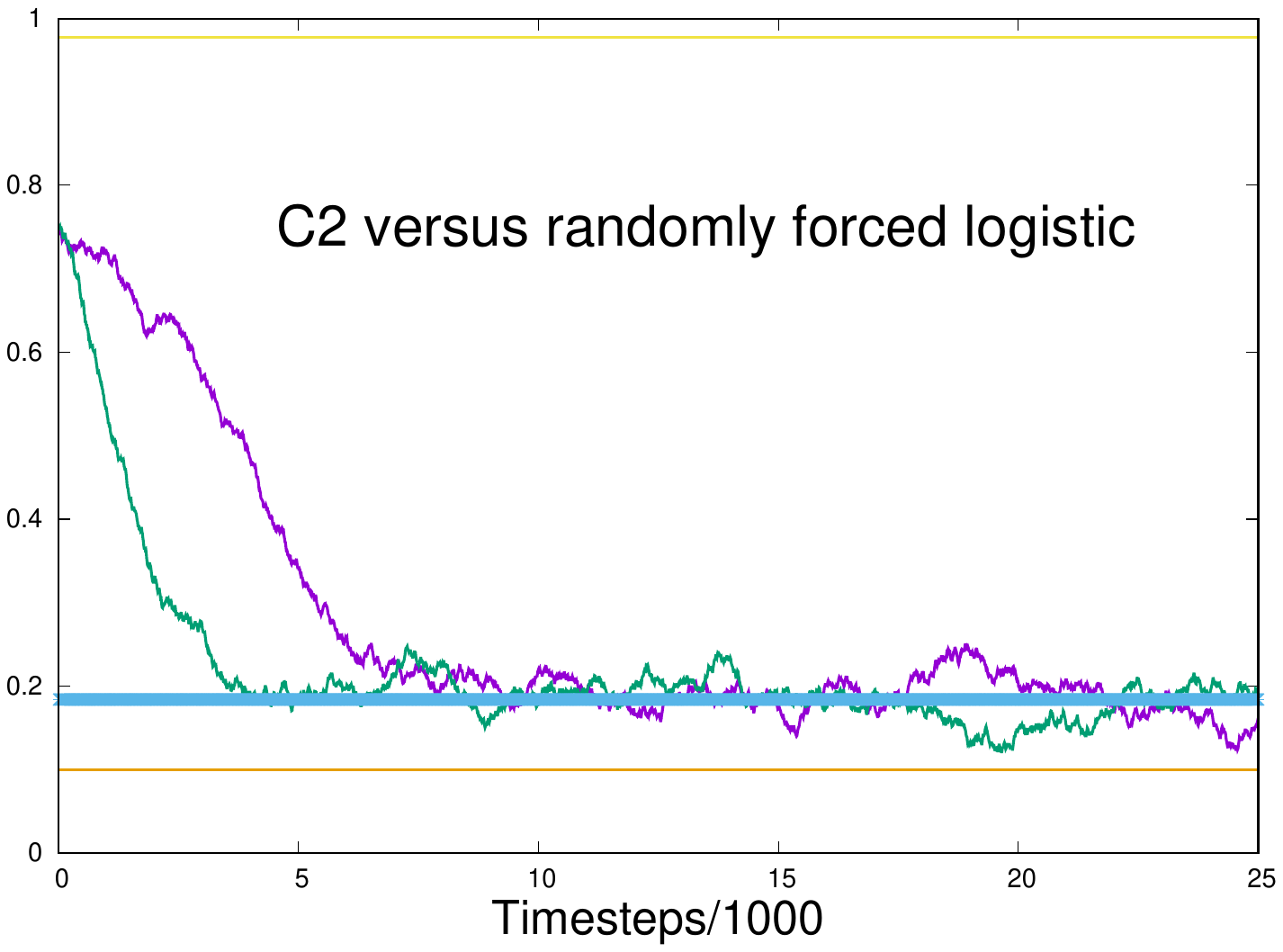}
\caption{The C2 solution $x_1(t)$ versus the randomly forced
logistic solution $x(t)$.
The main parameters are $a=b=1$, $x_0=0.75$, $f=0.15$, $f_r=0.05$.
The three horizontal lines represent (from top to bottom) the unstable, stable and 
linear steady state $f/a$. 
}
\label{fig:perturbed_logistic}
\end{figure}

In Fig.~\ref{fig:perturbed_logistic}, we show the C2 approximation in comparison with the logistic
solution, both obtained by a first order Euler time marching with $\Delta t=0.01$.
The other parameters are $a=b=1$, $x_0=0.75$, $f=0.15$ and $f_r = 0.05$.
As one can see, not only does C2 recover the correct statistical steady state to
a pretty good degree of accuracy but it also does it on a shorter time scale than the logistic dynamics. 
This is because the terms neglected by C2 would prolong the transient state, thereby 
deferring the attainment of the steady state. 
Precisely because of this, C2 acts as an efficient false transient technique.

\begin{figure}
\centering
\includegraphics[width=.9\linewidth]{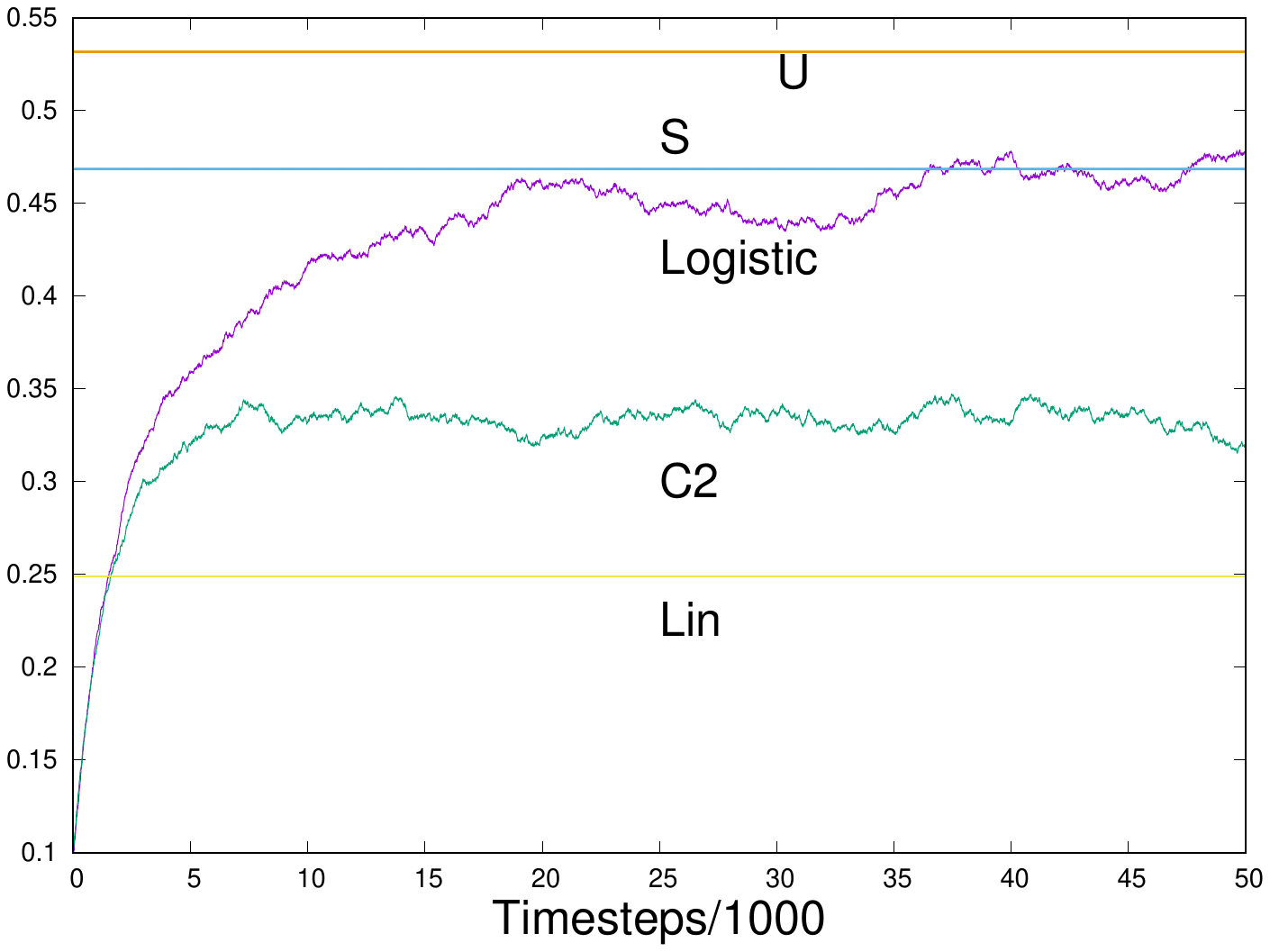}
\caption{The C2 solution $x_1(t)$ versus the randomly forced
logistic numerical solution $x(t)$.
The main parameters are $a=b=1$, $x_0=0.1$, $f=0.249$, $f_r=0.01$.
The three horizontal lines represent the unstable, stable and 
linear steady state $f/a$ (top to bottom). 
}
\label{fig:random_forced_log}
\end{figure}

In Fig.~\ref{fig:random_forced_log} we show a similar simulation, but with way more aggressive parameters,
$x_0=0.1, f=0.249$ and $f_r=0.01$, within three decimal places of the critical forcing.
As one can see, the C2 steady-state is now significantly less accurate.

\section{Auxiliary observations}
In the following, we comment on a few properties of the C2 approximation which may
prove useful for further developments of Carleman linearization of extended nonlinear
dynamical systems.

\subsection{Closure versus truncation}

It is natural to speculate that the accuracy of the Carleman approximation
would benefit by replacing truncation with a suitable linear closure.
In this case we illustrate the point quantitatively by replacing the
C2 truncation $x_3=0$ with a linear closure $x_3 = h x_2$ with $h$ an
adjustable parameter. Repeating the procedure discussed in the previous sections we
end up with the following C2 solution:
\begin{equation}
\label{CLOP}
x_1 = \frac{g^2}{1-g^2/(1-h)} \,.
\end{equation}
In the limit $h \to 0$ this reduces to the C2 truncation, whereas with 
$h=g^2$ it recovers the full C3 solution.
By equating (\ref{CLOP}) with the exact solution, one would readily
find out the precise expression $h=h(g)$ recovering the exact solution.
Clearly this is a circular argument, since it requires knowledge of
the solution itself. However, it shows that there is room 
for iterative procedures optimising the function $h=h(g)$.
More practically, it is readily shown that the simple position $h=g$
uplifts the generic $K-th$ Carleman approximant to order $K+1$.
This follows from the continued fraction expansion of the exact steady
state of the logistic equation.

\section{Growing logistic equation with forcing}

It is known that the Carleman linearization of the growing logistic faces
a divergence problem due to the finite radius of convergence
of the Carleman series. This, in turn, results from expanding the exact
solution in powers of $e^{\gamma t}$ instead of $e^{- \gamma t}$
which clearly diverges in the long-time limit.
In a recent paper, it was shown that the problem can be partially cured by
elegant regularization techniques based on analytic continuation~\cite{REG}.
In this Section, we show that unconditional stabilization of the growing logistic
can be obtained in a much simpler way by using a duality symmetry.

To this purpose, let us consider the growing forced logistic
in dimensionless form:
\begin{equation}
\label{GLE}
\dot x = x -x^2 - g^2 \,,
\end{equation}
where $-g^2$ now acts as a sink term for matters of symmetry.
The exact solution reads
\begin{equation}
x(t) = \frac{Sy(t)+U}{1+y(t)} \,,
\end{equation}
where $S$ and $U$ are swapped as compared to the decaying case, i.e.
$S=(1+\gamma)/2$, $U=(1-\gamma)/2$ and $\gamma=\sqrt{1-4g^2}$.

As is well known, Carleman linearization fails to capture the nonlinear saturation
of the exponential growth $x(t) = x_0 e^{\gamma t}$ in the initial transient. 
 A very simple and effective way of fixing the problem is to introduce
 the dual variable:  
 \begin{equation}
 \tilde x \equiv 1-x \,,
 \end{equation}
and show that it obeys the decaying logistic equation, with a 
sign change in the forcing term, namely 
\begin{equation}
\label{GLE2}
\dot {\tilde x} = - \tilde x +\tilde x ^2 + g^2 \,,
\end{equation}
which is exactly the decaying logistic with forcing discussed earlier on in this paper.
Hence, the failure of the Carleman linearization can be fully fixed by
simply moving from the original $x$ to its dual variable $\tilde x$.

\section{Fluid dynamic analogy}

The forced logistic equation can be loosely associated to a fluid dynamic
system with $a$ taking the role of dissipation, $b$ resembling the nonlinear inertia and
the forcing $f$ standing for the pressure gradient.
Within this analogy, the nonlinear parameter $g^2 = fb/a^2$ represents the
product of the ``Reynolds number'' $Re \sim b/a$ and $Re/Ma^2$
($\text{pressure}\times\text{gradient}/\text{dissipation}$) and $Ma$ standing for the Mach number~\cite{Cappelli_PoF}.
As a result, the identification is 
\begin{equation}\label{eq:g_scalar_definition}
g \leftrightarrow Re/Ma \equiv 1/Kn \,,
\end{equation} 
where $Kn$ is the Knudsen number.
In light of this analogy, one would conclude that the condition $g<1/2$ 
(the critical value above which the solution blows up) sets
a very stringent limit on the Reynolds numbers that can be treated by
Carleman linearization. 
Strictly speaking, the condition $Kn>1/2$ points to a rarefied gas regime
rather than fluids. In terms of Reynolds number, the constraint is
$Re < Ma/2$, well below $1$ for weakly compressible flows with $Ma \ll 1$. 
Of course, this analogy is only useful until it is not, because
the case of spatially extended systems involves a matrix problem 
at each time step rather than a simple scalar equation.
It is therefore expected that the spatial coupling may provide
a stabilizing mechanism, a point that we investigate later in the paper.

\subsection{Carleman embedding for spatially extended systems}

Let us generalize to the case of a scalar field $u(x,t)$ in $d=1$ 
spatial dimensions, namely the forced Burgers equation: 

\begin{equation}
\label{BUR}
\partial_t u + u \partial_x u = \nu \partial_x^2 u + f(x,t) \,,
\end{equation} 

In matrix form, after spatial discretization on a $N$-site lattice:

\begin{equation}
\label{BURMAT}
\dot u_i = -A_{ij} u_j + B_{ijk} u_j u_k + f_i\,,
\qquad
i,j,k=1,\dots,N
\end{equation}
where $A$ and $B$ are the lattice tensors associated to the discretized
diffusion and self-advection operators, respectively.

Let us write the equation~\eqref{BURMAT} in symbolic form as
\begin{equation} \label{eq:symbolic_u_eq}
\dot{U}_1 = -A_2 U_1 + B_3 U_2 + F_1 \,,
\end{equation}
where $\dot{U}_1 = \dot{u}_i, \,U_1 \equiv u_j$, $U_2 \equiv u_i u_j$, $F_1 \equiv f_i$,
$A_2 \equiv A_{ij}$ and $B_3 \equiv B_{ijk}$. 
In other words, the single subscript indicates the rank of the tensor. In this formalism, the
second order Carleman level writes as follows:
\begin{equation}
\dot{U}_2 = -A_4 U_2 + B_5 U_3 + F_3 U_1 \,,
\end{equation}
where the exact expressions of the symbols above are given in the Appendix~\ref{appendix}. 
By closing with $U_3 = 0$, the steady state of the above equation is given by:
\begin{equation}
U_2 = A_4^{-1} F_3 U_1 \,.
\end{equation}
Inserting this in the steady-state of~\eqref{eq:symbolic_u_eq}, we obtain:
\begin{equation}
U_1^{\mathrm{ss}}
=
\left(A_2 - B_3 A_4^{-1} F_3\right)^{-1} F_1 \,.
\end{equation}
That is
\begin{equation}
U_1^{\mathrm{ss}}
=
\left(I - A_2^{-1} B_3 A_4^{-1} F_3\right)^{-1}
A_2^{-1} F_1 \,.
\end{equation}
This is the matrix analogue of the scalar equation~\eqref{eq:g_scalar_definition} with
\begin{equation}\label{eq:g2_matrix_definition}
G_2 = A_2^{-1} B_3 A_4^{-1} F_3 \,.
\end{equation}
Note that $G_2$ is a $G=N \times N$ rank-2 matrix. Clearly, the condition
\begin{equation}
\lVert G_2 \rVert < 1 \,,
\end{equation}
sets a limit to the convergence of the $C_2$ approximation. 
Translating such condition into a corresponding critical value of the Reynolds 
number, as per the analogy discussed in the previous Section, is not 
straightforward, and we defer this task to future work. 
In the following, we provide instead some numerical results using the 
Carleman lattice Boltzmann method.

\subsection{Carleman Lattice Boltzmann}

Before delving into details regarding the numerical implementation of lattice Boltzmann, it is appropriate 
to introduce its core ideas and the second--order Carleman implementation.
\\
Consider the lattice Boltzmann Method (LBM), specifically a forced D2Q9 
model~\cite{Succi2001, BENZI1992} (nine discrete distributions in two spatial dimensions). 
In LBM, the fluid is described by a discrete set of distribution functions, or populations $f_i$, each 
associated with the discrete lattice velocities $\mathbf{c_x}_i = (c_{xi}, c_{yi})$. 
Each component $\mathbf{c}_i$ can take the values $-1$, $0$, or $1$, which 
in the D2Q9 model gives rise to a total of $Q=9$ distributions. 
\\
The dynamics is governed by the interplay of two computational steps: collision and advection.
The collision, or relaxation, step is local but nonlinear and is given by
\begin{equation}
\label{eq:collision}
f^*_i(\mathbf{x},t)
=
(1-\omega)\,f_i(\mathbf{x},t)
+\omega\,f_i^{\mathrm{eq}}(\mathbf{x},t)
+F_i(\mathbf{x}) \,,
\end{equation}
where $F_i(\mathbf{x})$ represents an external forcing term, $f_i^{\mathrm{eq}}(\mathbf{x},t)$ is the equilibrium distribution, and $\omega$ is the relaxation parameter. 
The latter is related to the kinematic viscosity through
\begin{equation}
\label{eq:viscosity}
\nu
=
c_s^2\left(\frac{1}{\omega}-\frac{1}{2}\right),
\end{equation}
$c_s$ being the lattice speed of sound. 
The local equilibrium distribution contains the nonlinear dependence 
on the fluid velocity and is given by
\begin{equation}
\label{eq:equilibrium}
f_i^{\mathrm{eq}}(\mathbf{x},t)
=
w_i\rho(\mathbf{x},t)
\left(
1
+\frac{\mathbf{u}\cdot\mathbf{c}_i}{c_s^2}
+\frac{(\mathbf{u}\cdot\mathbf{c}_i)^2}{2c_s^4}
-\frac{\mathbf{u}\cdot\mathbf{u}}{2c_s^2}
\right),
\end{equation}
where $w_i$ are the lattice weights associated with the different populations, while 
$\mathbf{u}$ denotes the local fluid velocity.

The advection, or streaming, step is instead linear but nonlocal, and reads
\begin{equation}
\label{eq:streaming}
f_i(\mathbf{x}+\mathbf{c}_i\Delta t,t+\Delta t)
=
f_i^*(\mathbf{x},t).
\end{equation}

The macroscopic density and current density are obtained from the zeroth- and 
first-order moments of the populations,
\begin{equation}\label{eq:observables}
\rho = \sum_{i=1}^Q f_i,
\qquad
\mathbf{J} = \sum_{i=1}^Q f_i \,\mathbf{c}_i = \rho\,\mathbf{u}.
\end{equation}
Substituting these expressions into Eq.~\eqref{eq:equilibrium}, the collision step can be written explicitly as a function of the populations $f_i$. For incompressible or weakly compressible flows, characterized by a small Mach number, $Ma=|\mathbf{u}|/c_s\ll1$, Eq.~\eqref{eq:collision} can be expressed in the generic polynomial form
\begin{equation}
\label{eq:tensor_collision}
f^*_i
=
A_{ij} \, f_j
+B_{ijk} \, f_jf_k
+C_{ijkl} \, f_jf_kf_l
+F_i\, ,
\end{equation}
where the third-order term originates from the approximation $1/\rho\approx2-\rho$, which is valid for $\rho\approx1$ in the weakly compressible regime.
The explicit expressions of the tensors $A$, $B$, and $C$ can be found in~\cite{sanavio_CLB}. For compactness, the space-time dependence has been omitted, since the collision step is local. 
This representation explicitly exposes the polynomial nonlinear structure of the 
LBM collision operator and provides the natural starting point for its Carleman linearization.

The incompressibility constraint also defines a practical range of validity for the LBM description of fluid flows, since the correspondence with the Navier--Stokes equations is obtained only in the low-Mach-number regime. In the standard lattice formulation, the speed of sound $c_s$ is fixed, so this condition translates directly into a constraint on the magnitude of the fluid velocity, and hence of the current. For $c_s^2=1/3$ and $\rho\simeq1$, we therefore expect the incompressible regime to require $|\mathbf{J}|\ll c_s$, with $|\mathbf{J}| \lesssim 0.1$ providing a conservative practical bound.

We can now apply the Carleman linearization to the LBM dynamics. Introducing the second-order lifted variables
\begin{equation}
g_{ij}(\mathbf{x}_1,\mathbf{x}_2)
=
f_i(\mathbf{x}_1)f_j(\mathbf{x}_2),
\end{equation}
the nonlinear collision operator is mapped onto a hierarchy of equations involving increasingly high-order products of the populations. Truncating this hierarchy at second order yields the Carleman lattice Boltzmann (CLB) approximation, denoted by C2. In compact notation, the resulting collision step takes the form
\begin{align}\label{eq:C1}
f_i^* &= A_{ij}\,f_j + B_{ijk}\,g_{jk} + F_i,
\\
\label{eq:C2}
g_{ij}^* &= A_{ik}A_{jl}\, g_{kl} 
+ A_{ik}f_k\,F_j + F_j\,A_{ik}f_k 
+ F_i\,F_j 
\\ & \nonumber
+ B_{ikl}g_{kl}\,F_j + F_i\,B_{jkl}g_{kl}  \, .
\end{align}
This set of equations has a structure analogous to that used in~\cite{Cappelli_PoF}.
This formulation allows us to test whether the instability observed in the truncated Carleman dynamics of the logistic equation persists in the substantially richer setting of forced LBM.
\\
We therefore investigate the C2 dynamics numerically and compare it with the corresponding LBM solution. We focus on the evolution of the current and density fields, and in particular on whether the truncated dynamics remains within the low-Mach-number regime or develops the runaway behavior suggested by the logistic analogy.

\FloatBarrier
\section{Numerical results}

It is difficult to establish a direct correspondence between the condition $\lVert G_2\rVert<1$ and the critical Reynolds number above which C2 convergence is lost. In the following, we investigate this question by means of direct numerical simulations based on the Carleman lattice Boltzmann method.

The numerical tests were performed on a $64\times64$ grid to verify whether C2 could correctly reproduce the stationary state in regimes with low and mild Reynolds numbers.
\\
Defining the Reynolds number as the ratio between inertial and dissipative forces
\begin{equation}\label{eq:reynolds}
Re=\frac{U \,N}{2\nu}\,, \qquad
U = \max_{\{\mathbf{x},t\}}
\left(
|u_x(\mathbf{x},t)|, |u_y(\mathbf{x},t)|
\right) \,,
\end{equation}
we identify the \textit{low--Reynolds} regime as $1\leq Re<10$ and the \textit{mild-Reynolds} regime as $10\leq Re<100$.

\begin{figure*}
    \centering
    \includegraphics[width=\linewidth]{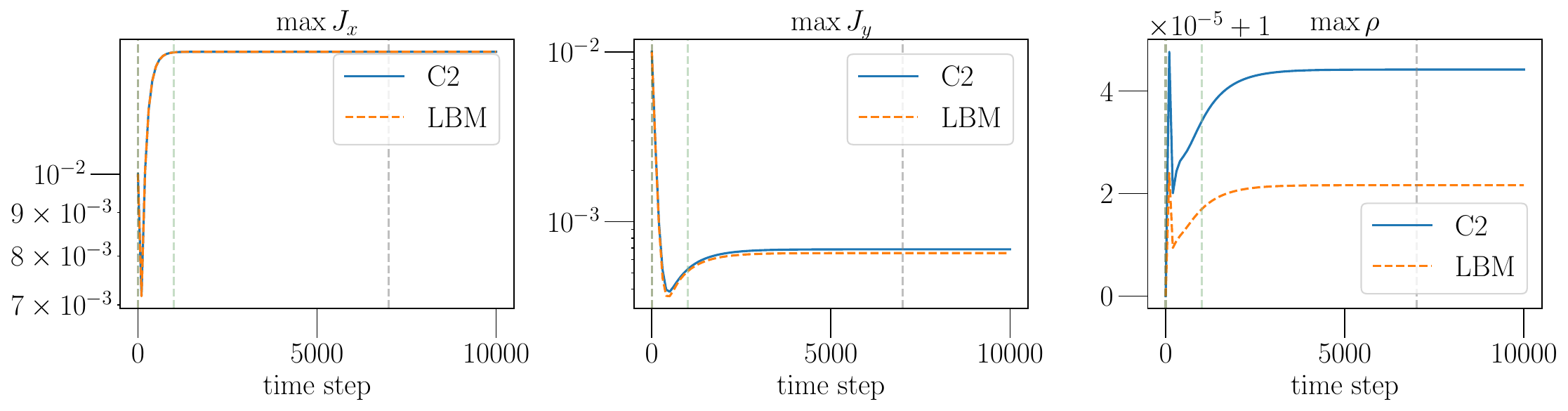}
    \caption{
    From left to right: maximum value of the current along the two spatial axes and maximum value of the density $\rho$, obtained from LBM and C2 simulations of the Kolmogorov-like flow at $Re\approx3$, with $\omega=1$. C2 correctly reproduces the steady-state flow, remaining within the low-Mach-number and weakly compressible regime, with density fluctuations of $\mathcal{O}(10^{-5})$. The vertical dashed lines indicate the time frames chosen to represent the flow fields in Fig.~\ref{fig:fields_low_Re}.
    }
    \label{fig:j_and_rho_low_Re}
\end{figure*}

\begin{figure*}
    \centering
    \includegraphics[width=\linewidth]{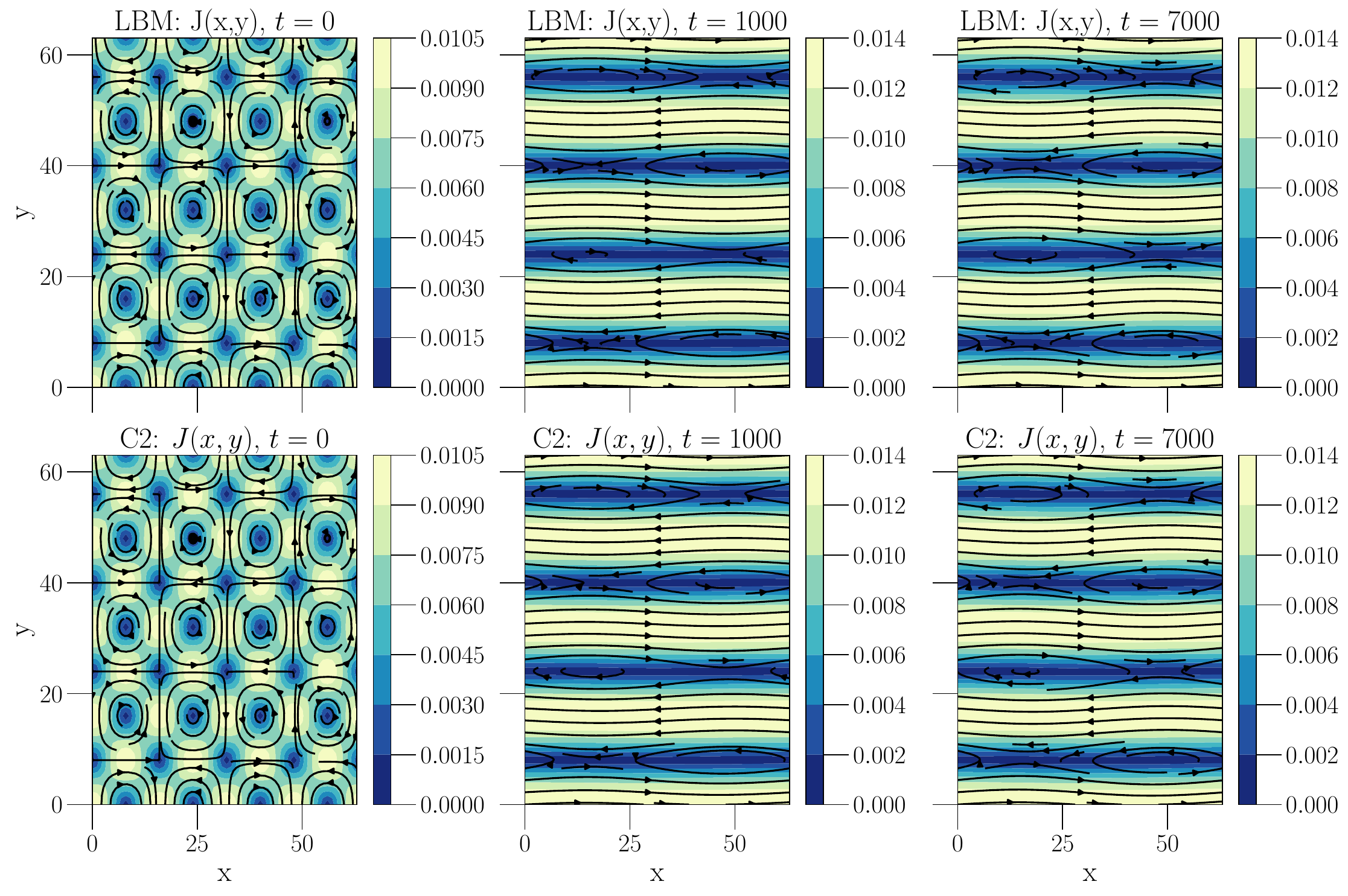}
    \caption{
    Magnitude and streamlines of the current $\mathbf{J}$ at different times for the LBM (first row) and CLB (second row) simulations with $\omega=1$, corresponding to $Re\approx3$. The first column shows the two initial configurations, while the central column corresponds to a time immediately after the initial transient. The final column shows the two configurations at $t=7000$, when the flow has reached a steady state. The selected times are those indicated by the vertical dashed lines in Fig.~\ref{fig:j_and_rho_low_Re}. In this low-Reynolds-number regime, C2 correctly reproduces the steady state obtained with LBM.
    }
    \label{fig:fields_low_Re}
\end{figure*}

We considered an initial condition that already contains a symmetric vortical structure and coupled it with a Kolmogorov--like forcing to trigger the formation of additional vortices.
\\
The initial distribution is obtained by inserting the starting velocity configuration into Eq.~\eqref{eq:equilibrium} 
\begin{equation}
\label{eq:initial_velocity}
\begin{split}
    &u_x(x,y) = u_0 \sin(kx) \sin(ky), 
    \\ &
    u_y(x,y) = u_0 \cos(kx) \cos(ky),    
\end{split}
\end{equation}
where $k=4\pi/N$ and $u_0=0.01$.
Similarly, we introduced the external forcing term
\begin{equation}\label{eq:forcing}
F_i(x,y) = \frac{w_i}{c_s^2}  \left(
{F_0}_x\cos({k_F}_x\,y) \,{c_x}_i
+
{F_0}_y\cos({k_F}_y\,x) \,{c_y}_i
\right)\,,
\end{equation}
with $F_0=(9\times10^{-5}, \,10^{-6})$ and $k_F=(4\pi/N, \, 2\pi/N)$.
Therefore, energy is injected into the system mainly along the $x$ direction, exciting the second-largest mode, while the component along the $y$ direction prevents the corresponding velocity component from decaying to zero. In this way, the nonlinear contribution remains significant along both components, provided that it is not damped out by excessive viscosity.
We first investigate the behavior of C2 and LBM in a low-Reynolds regime by choosing a comparatively high value of the numerical viscosity, $\nu=1/6$, corresponding to $Re\approx3$. 
This is obtained by setting the relaxation parameter to $\omega=1$.
In this scenario, we expect the viscous effects to dominate the nonlinear contributions and prevent the formation of more complex vortical structures. This is indeed observed in Figs.~\ref{fig:j_and_rho_low_Re} and~\ref{fig:fields_low_Re}.
In Fig.~\ref{fig:j_and_rho_low_Re}, the maximum values of the current $\mathbf{J}$ and the density are shown, computed as in Eq.~\eqref{eq:observables}. In this low-Reynolds-number regime, after an initial transient, the external forcing stabilizes the values of $\mathbf{J}$ which remain constant over approximately $10^4$ time steps. 
We point out that, thanks to the forcing along the $y$ axis, $J_y$ does not decay to zero. In principle, this should contribute to the development of nonlinear structures within the flow, but, as anticipated, these effects are damped by the viscosity. Nevertheless, during this interval, C2 correctly reproduces the values obtained with LBM.
\\
Three representative time frames of the evolution are also highlighted by vertical dashed lines: the initial configuration, the state immediately after the transient, and the configuration after a long evolution time. Indeed, the flow configuration is not necessarily stationary even when the maximum values of the observables remain constant. For this reason, we compare the current field $\mathbf{J}$ and its streamlines at the aforementioned time frames. These are shown in Fig.~\ref{fig:fields_low_Re}, where it is evident that the current $\mathbf{J}$ remains almost unchanged between the state immediately after the transient and the final stage of the evolution.
Specifically, the system aligns with the dominant component of the external forcing, reproducing a quasi-laminar one-dimensional flow. Only small perturbations are present, indicating a weak nonlinear contribution. This suggests that the nonlinear effects that would lead to the formation of vortices are counteracted by the viscous term before the vortical structures can fully develop, resulting in a quasi-laminar stationary state.
\\
In this low-Reynolds-number regime, where no vortices or more complex structures develop, we can conclude that C2 is able to correctly reproduce the steady state obtained with LBM.

If we consider the convergence condition derived for the logistic equation and reported in Eq.~\eqref{eq:g_scalar_definition}, we observe that the ratio $Re/Ma\approx111$ significantly exceeds the critical bound imposed by the logistic case. This difference is likely due to the more complex interactions present in LBM, which appear to soften this constraint, most likely due to diffusive effects~\cite{Cappelli_PoF}.
\\
However, no rigorous bound relating the effectiveness of C2 to the Reynolds number is currently available. We therefore proceed by increasing the complexity of the flow and the Reynolds number, with the aim of empirically identifying the limit of the C2 approximation.
\\

In the previous case, the relatively high viscosity prevented the growth of more complex nonlinear structures. To overcome this limitation, we performed a second set of simulations, keeping the same initial conditions but setting $\omega=1.5$, corresponding to a numerical viscosity of $\nu=1/18$, and evolving the system for $10^4$ time steps until the flow appeared to reach a stationary state.
\\
For this simulation, the corresponding Reynolds number is approximately $24$, placing the system well within the mild-Reynolds regime.

\begin{figure*}
    \centering
    \includegraphics[width=\linewidth]{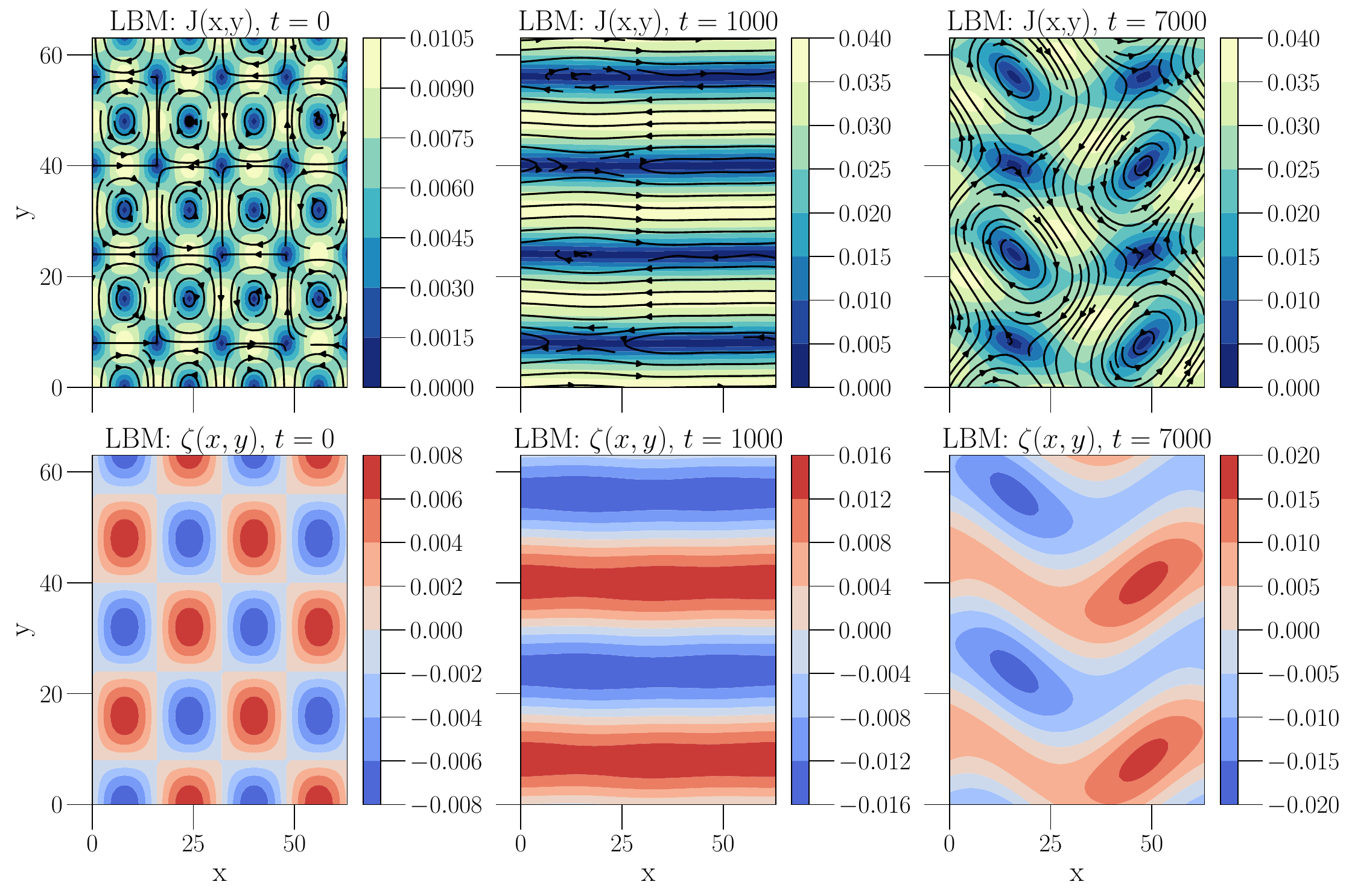}
    \caption{
        Current density $\mathbf{J}$ (top row) and vorticity (bottom row) at different times for the LBM solution on a $64\times 64$ grid with $\omega=1.5$, which corresponds to $ Re \approx 24$.
        From left to right, the panels show the initial state ($t=0$), an approximately linear regime ($t=1000$), and the final stationary vortical state ($t=7000$), characterized by a time-independent vortex configuration. In the top row, the streamlines illustrate the flow structure, while the color scale represents the magnitude of $\mathbf{J}$. The vorticity $\mathbf{\zeta}$ is shown in the bottom row using a distinct color scale.
    }
    \label{fig:field_and_vorticity_LBM}
\end{figure*}

In Fig.~\ref{fig:field_and_vorticity_LBM}, the flow configuration and vorticity are shown at the three stages of the system's evolution. As anticipated, at the initial time the flow already exhibits a highly symmetric vortical structure, resulting from the particular choice of the initial conditions. Subsequently, under the action of the forcing, the system develops a quasi-laminar structure that resembles the form of $F_x={F_0}_x \cos({k_F}_x\,y)$. Small perturbations can already be observed within this flow, eventually leading to the formation of staggered vortical structures. These become clearly visible at a later stage of the evolution. We chose to show the configuration at time $t=7000$, when the vortical pattern has already formed, although it remains essentially frozen during the final stages of the evolution.

Fig.~\ref{fig:placeholder} shows the evolution of the maximum and minimum values of the current $\mathbf{J}$, highlighting the three distinct stages of the evolution. Notice that the $y$ component does not decay to zero but is maintained by the forcing. This allows the vortical structures to form through the nonlinear term, which extracts energy from $J_x$ and transfers it to $J_y$, and then back to $J_x$.

\begin{figure*}
    \centering
    \includegraphics[width=\linewidth]{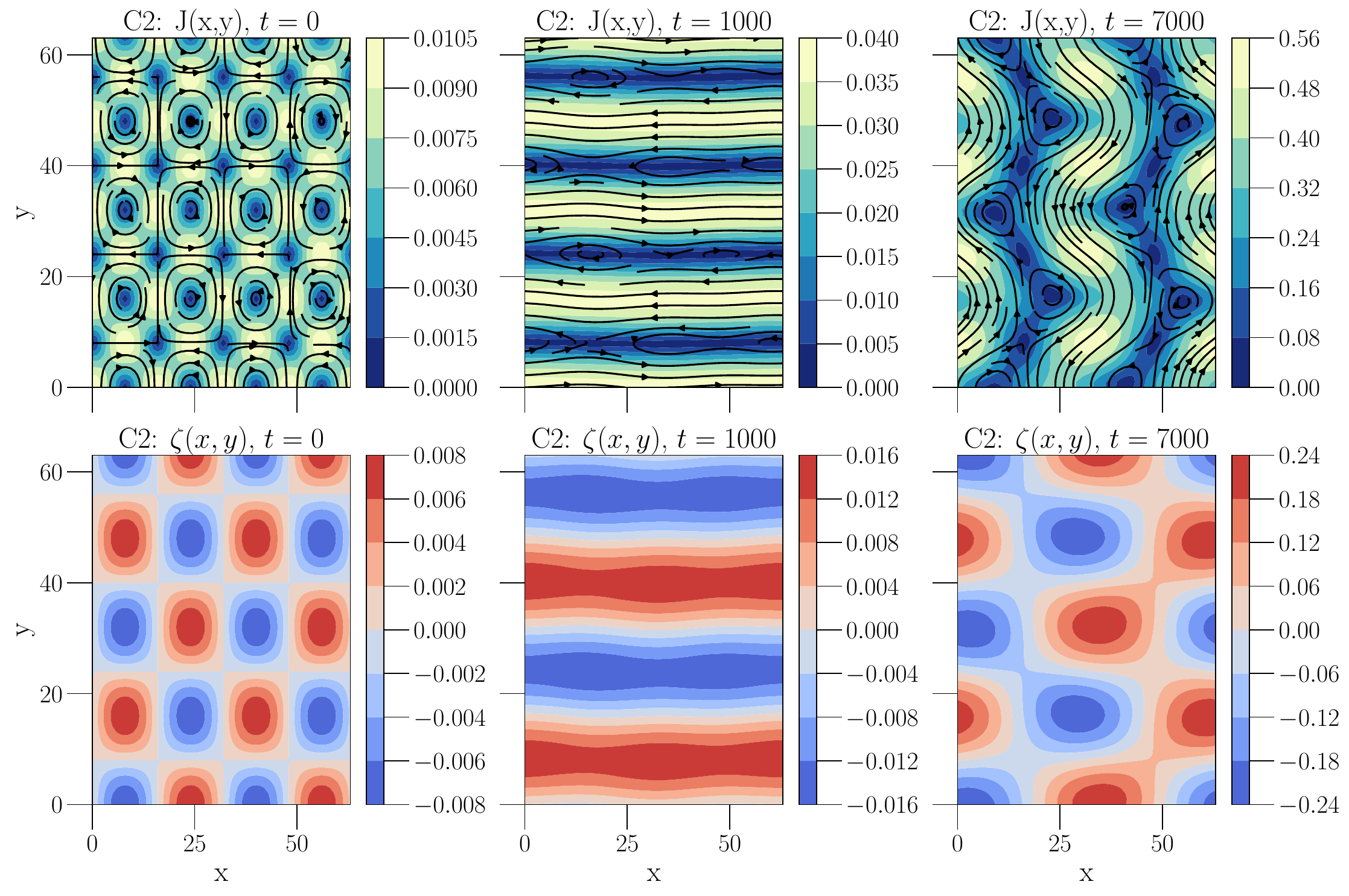}
    \caption{
    Analogously to Fig.~\ref{fig:field_and_vorticity_LBM}, the current density $\mathbf{J}$ (top row) and vorticity $\mathbf{\zeta}$ (bottom row) are shown at different times for the second-order Carleman (C2) solution. From left to right, the panels correspond to the initial state ($t=0$), an approximately linear regime ($t=1000$), and the late-time state ($t=7000$). While the initial evolution is consistent with the LBM solution, C2 fails to recover the stable vortical configuration observed at late times. Instead, the current grows progressively and eventually exceeds the bound imposed by the incompressibility condition. Moreover, the resulting vortex structures differ qualitatively from those obtained with LBM, indicating that the C2 approximation does not correctly capture the nonlinear dynamics leading to the stable vortical state. As in Fig.~\ref{fig:field_and_vorticity_LBM}, streamlines are shown in the top row, with the color scale representing the magnitude of $\mathbf{J}$, while the vorticity is displayed in the bottom row using a distinct color scale.
    }
    \label{fig:field_and_vorticity_C2}
\end{figure*}

\begin{figure*}
    \centering
    \includegraphics[width=\linewidth]{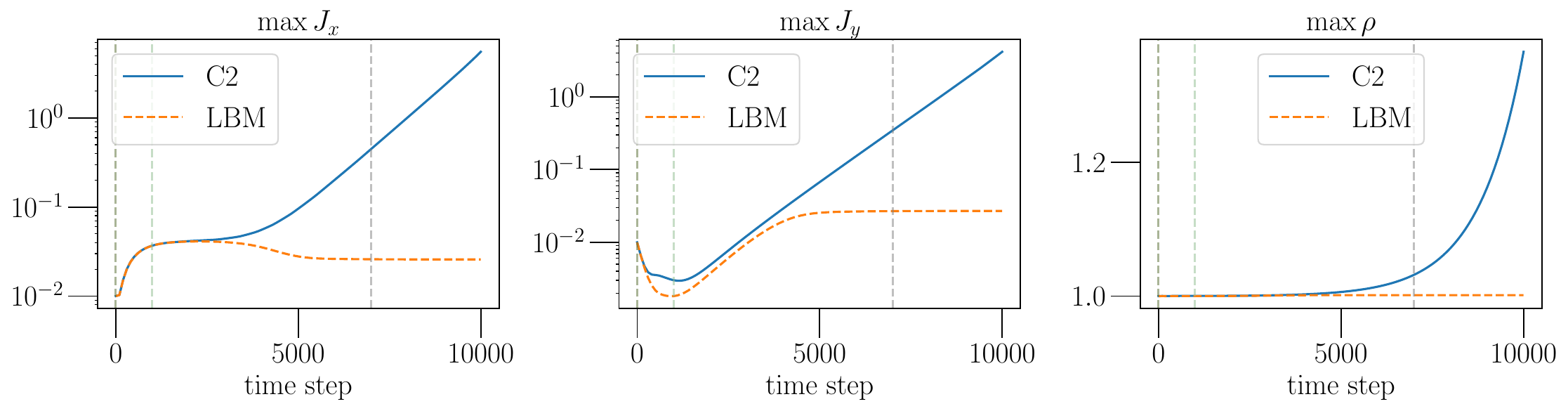}
    \caption{
    From left to right: maximum values of the current density $\mathbf{J}$, and of the density field $\rho$ for the LBM solution (dashed line) and the second-order truncated Carleman approximation (C2). The results correspond to the initial conditions given in Eq.~\eqref{eq:initial_velocity}, with $\omega=1.5$ and $Re\approx 24$. The vertical dashed lines indicate the time frames shown in Figs.~\ref{fig:field_and_vorticity_LBM} and~\ref{fig:field_and_vorticity_C2}. The figure highlights the progressive departure of the C2 approximation from the LBM solution, providing evidence that this regime is analogous to the case $g^2>1/4$ in the logistic model. 
    }
    \label{fig:placeholder}
\end{figure*}

To verify whether C2 could reproduce the vortical stationary state, we evolved the same initial conditions using Eqs.~\ref{eq:C1} and~\eqref{eq:C2}. Fig.~\ref{fig:field_and_vorticity_C2} shows the configuration of the current $\mathbf{J}$ and the scalar vorticity $\zeta$, while Fig.~\ref{fig:placeholder} compares the maximum values of $\mathbf{J}$ and the density $\rho$ with those obtained using LBM. The two figures, together with Fig.~\ref{fig:field_and_vorticity_LBM}, show that C2 correctly reproduces the initial quasi-linear phase of the system but fails to reproduce the subsequent formation of vortices. Instead, the system appears to diverge completely, similarly to what is usually observed in the case of the growing logistic.
\\
This makes physical sense because vortex formation is driven by local instabilities.

Indeed, both the current and the density exhibit an increasing trend that exceeds the validity range of LBM in the low-Mach-number and incompressible regimes. This suggests that the observed divergence is related to the validity range of C2.
As in the previous case, the specific value of $\lVert G\rVert$ is not directly available, but the ratio $Re/Ma$ is approximately $333$.
Specifically, the consistent growth behaviour suggests that the issue has a similar origin to the one observed in the growing logistic case, where a change of variables proves able to fix the problem. 
However, at present, it is not clear how to generalize this strategy to the CLB framework because the exact steady state is unkown. 

\section{Summary}

We have discussed some basic time-asymptotic properties of the Carleman linearization 
of the decaying forced logistic equation and shown that the second-order Carleman 
approximation captures the steady state with second-order accuracy in the forcing parameter. 
In this context, the convergence threshold can be related to a quantity analogous 
to the Reynolds-to-Mach-number ratio.

We have also shown that a suitable closure can extend the convergence 
horizon to third-order Carleman truncation. Furthermore, a suitable change of variables 
can stabilize the case in which the Carleman approximation exhibits exponential growth 
in the initial stage of the evolution. 

The logistic equation thus reveals a number of interesting analytical properties which may bear
on the convergence of Carleman approximations of the fluid equations. 
We then sought to determine to what extent these insights can be extended to 
more complex fluid systems. To this end, we generalized the convergence criterion for the scalar case
to the multidimensional context, formally obtaining a corresponding  matrix relation.
Such relation could be inspected by means of spectral analysis techniques, a task
that we defer to future investigations.
We proceeded instead by means of numerical simulations of a two-dimensional flow 
on a $64\times64$ lattice, using a high-performance multi-GPU version of the CLB2 code.
We note that multiple GPUs are required because of the large memory footprint of the CLBs code,
over $10$ Gbytes for a $64^2$ grid C2 simulation. 

Specifically we focused on the lattice Boltzmann framework, comparing the results obtained from LBM and C2 with two different viscosity values, corresponding to a low-Reynolds regime with $Re\approx3$ and to a mild-Reynolds regime with $Re\approx24$. At low Reynolds number, C2 correctly reproduces the stationary state obtained with LBM, while at mild Reynolds number, it fails to reproduce the nonlinear evolution leading to the formation of static vortical structures. In the latter case, C2 eventually diverges, showing untamed growth in time very reminiscent of the 
C2 approximation of the growing logistic equation.

Interestingly, the numerical results do not show a one-to-one correspondence with the 
bound imposed by the logistic equation. In fact, they exceed this bound by 
about two orders of magnitude, which we interpret as a stabilisation effect due to diffusive coupling. 
Nevertheless, there still is an upper bound on the validity of C2: at low Reynolds 
numbers, C2 is able to capture the stationary state, while at mild Reynolds numbers with more 
complex dynamics it exhibits divergent behaviour reminiscent of the growing logistic case.
\\
Thus, even though there is no one-to-one quantitative correspondence 
between the two systems, a qualitative analogy appears to exist, shedding some 
hope that the techniques developed for the logistic equation 
might carry over to CLB. Specifically, closing the Carleman hierarchy instead of truncating
it might serve as a starting point to develop a new class of regularization
strategies for the CLB simulation of higher Reynolds numbers.
Likewise, suitable extensions of the duality property of the forced
logistics may soften the observed runaway behaviour.
These topics make interesting subjects for future work.

\FloatBarrier
\section*{Acknowledgments}
The authors acknowledge valuable discussions with M. M\"oller, 
M. Lacatus and C. Sanavio.
The authors have no conflicts to disclose. 
The data that support the findings of this study are available from the corresponding 
author upon reasonable request.

\appendix

\section{C2 Matrices in coordinate form} \label{appendix}

Let us start with two replicas of the original equation 
\begin{eqnarray}
\dot u_i = A_{ij} u_j + B_{ijk} u_j u_k + f_i \,,\\
\dot u_l = A_{lm} u_m + B_{lmn} u_m u_n + f_l \,.
\end{eqnarray}
Multiplying the first by $u_l$ and the second by $u_i$ we obtain
\begin{equation}
\begin{split}
\frac{d u_i u_l}{dt} = 
&
A_{ij} u_j u_l + A_{lm} u_m u_i + 
+ f_i u_l + f_l u_i
\\&
A_{ij} B_{lmn} u_j u_m u_n + A_{lm} B_{ijk} u_m u_j u_k \,.
\end{split}
\end{equation}
The first defines the $A_4$ matrix:
\begin{equation}
    A_{4,ijlm} = A_{ij}\delta_{lm} + A_{lm} \delta_{ij}    \,,
\end{equation}
as applied to $U_2 \equiv u_j u_m$.
This is the matrix analogue of the coefficient $2a$ for the scalar case.
Likewise
\begin{equation}
F_{3,ilm} = f_i \delta_{lm} + f_l \delta_{im} \,,    
\end{equation}
as applied to $u_m$.
The tensor $B_5$ is obtained in a similar way, but we omit it
because it plays no role at the C2 level.

\bibliography{biblio}

\end{document}